\documentclass[conference]{IEEEtran}
\IEEEoverridecommandlockouts
\usepackage{cite}
\usepackage{amsmath,amssymb,amsfonts}
\usepackage{algorithmic}
\usepackage{graphicx}
\usepackage{textcomp}
\usepackage{listings}
\usepackage{xcolor}
\usepackage{url}
\usepackage{subfig}

\def\BibTeX{{\rm B\kern-.05em{\sc i\kern-.025em b}\kern-.08em
    T\kern-.1667em\lower.7ex\hbox{E}\kern-.125emX}}

\colorlet{punct}{red!60!black}
\definecolor{background}{HTML}{EEEEEE}
\definecolor{delim}{RGB}{20,105,176}
\colorlet{numb}{magenta!60!black}

\lstdefinelanguage{json}{
    basicstyle=\fontsize{7}{8}\selectfont\ttfamily,
		numbers=right, 
		numberstyle=\footnotesize, 
		numbersep=8pt, 
    showstringspaces=false,
    breaklines=true,
		float=tp,
		floatplacement=tbp,
		belowskip=-20pt,
		xleftmargin=\parindent,
    literate=
     *{0}{{{\color{numb}0}}}{1}
      {1}{{{\color{numb}1}}}{1}
      {2}{{{\color{numb}2}}}{1}
      {3}{{{\color{numb}3}}}{1}
      {4}{{{\color{numb}4}}}{1}
      {5}{{{\color{numb}5}}}{1}
      {6}{{{\color{numb}6}}}{1}
      {7}{{{\color{numb}7}}}{1}
      {8}{{{\color{numb}8}}}{1}
      {9}{{{\color{numb}9}}}{1}
      {:}{{{\color{punct}{:}}}}{1}
      {,}{{{\color{punct}{,}}}}{1}
      {\{}{{{\color{delim}{\{}}}}{1}
      {\}}{{{\color{delim}{\}}}}}{1}
      {[}{{{\color{delim}{[}}}}{1}
      {]}{{{\color{delim}{]}}}}{1},
}

\lstdefinelanguage{jsonnonumber}{
    basicstyle=\footnotesize\ttfamily,
    showstringspaces=false,
    breaklines=true,
		xleftmargin=\parindent,
    literate=
     *{0}{{{\color{numb}0}}}{1}
      {1}{{{\color{numb}1}}}{1}
      {2}{{{\color{numb}2}}}{1}
      {3}{{{\color{numb}3}}}{1}
      {4}{{{\color{numb}4}}}{1}
      {5}{{{\color{numb}5}}}{1}
      {6}{{{\color{numb}6}}}{1}
      {7}{{{\color{numb}7}}}{1}
      {8}{{{\color{numb}8}}}{1}
      {9}{{{\color{numb}9}}}{1}
      {:}{{{\color{punct}{:}}}}{1}
      {,}{{{\color{punct}{,}}}}{1}
      {\{}{{{\color{delim}{\{}}}}{1}
      {\}}{{{\color{delim}{\}}}}}{1}
      {[}{{{\color{delim}{[}}}}{1}
      {]}{{{\color{delim}{]}}}}{1},
}

\lstdefinelanguage{jsonnonumbersmall}{
  basicstyle=\fontsize{6}{7}\selectfont\ttfamily,
    showstringspaces=false,
    breaklines=true,
		xleftmargin=\parindent,
    literate=
     *{0}{{{\color{numb}0}}}{1}
      {1}{{{\color{numb}1}}}{1}
      {2}{{{\color{numb}2}}}{1}
      {3}{{{\color{numb}3}}}{1}
      {4}{{{\color{numb}4}}}{1}
      {5}{{{\color{numb}5}}}{1}
      {6}{{{\color{numb}6}}}{1}
      {7}{{{\color{numb}7}}}{1}
      {8}{{{\color{numb}8}}}{1}
      {9}{{{\color{numb}9}}}{1}
      {:}{{{\color{punct}{:}}}}{1}
      {,}{{{\color{punct}{,}}}}{1}
      {\{}{{{\color{delim}{\{}}}}{1}
      {\}}{{{\color{delim}{\}}}}}{1}
      {[}{{{\color{delim}{[}}}}{1}
      {]}{{{\color{delim}{]}}}}{1},
}

\definecolor{pblue}{rgb}{0.13,0.13,1}
\definecolor{pgreen}{rgb}{0,0.5,0}
\definecolor{pred}{rgb}{0.9,0,0}
\definecolor{pgrey}{rgb}{0.46,0.45,0.48}

\newcommand{\egeon}{\textsc{Egeon}}

\begin{document}
\bstctlcite{IEEEexample:BSTcontrol}

\title{\egeon: Software-Defined Data Protection for Object Storage}

\author{\IEEEauthorblockN{Raul Saiz-Laudo, Marc S\'{a}nchez-Artigas}
\IEEEauthorblockA{\textit{Computer Science and Maths, Universitat Rovira i Virgili} \\
Tarragona, Catalonia, Spain}
}

\maketitle

\begin{abstract}
With the growth in popularity of cloud computing, object storage  systems (e.g., Amazon S$3$,  OpenStack Swift, Ceph)
have gained momentum for their relatively low per-GB costs~and high availability. However, as increasingly more sensitive 
data is being accrued, the need to natively integrate privacy controls~into the storage is growing in relevance. Today, 
due to the poor object storage interface, privacy controls are enforced by data curators
with full access to data in the clear. This motivates the need for a new approach to data privacy that can provide strong assurance
and control to data owners. To fulfill this need, this paper presents \egeon, a novel software-defined data protection framework for object storage. 
\egeon~enables users to declaratively set privacy policies on how their data can be shared. In the privacy policies, the users can
 build complex data protection services~through~the composition of data transformations, which are invoked inline~by \egeon~upon a read request. 
As a result, data owners can trivially display multiple views from the same data piece, and modify these views by only updating the policies.
And all without restructuring the internals of the underlying object storage system. The  \egeon\hphantom{} prototype has  been  built atop
OpenStack Swift. Evaluation results shows promise in developing data
protection services with little overhead directly into the object store. 
Further, depending on~the amount of data filtered out in the transformed views, end-to-end
latency can be low due to the savings in network communication.
\end{abstract}


\section{Introduction}

With the rapid growth in popularity of Cloud services, object storage  systems (e.g.,
 Amazon S$3$, IBM COS  or OpenStack Swift~\cite{swift}) have gained momentum.
These storage systems offer consolidated storage at scale, with high degrees of availability 
and bandwidth at low cost. Proof of that is the recent trend of serverless computing. 
Due to the high difficulty of function-to-function communication\footnote{Some
works have shown that cloud functions can communicate directly using NAT (Network Address Translation) traversal techniques. 
However, direct communication between functions is not supported by cloud providers.}, 
many serverless systems use
object storage for passing data between functions~\cite{Sampe2018, pu2019shuffling, muller20, primula, sonic},
which has revived the interest in this type of object storage. 

~Although very useful for cloud applications, object storage systems offer
a small number of options to keep sensitive data safe. Few (or no) efforts 
have been realized on security issues such as data confidentiality, data integrity, or access control, to mention a few. 
For instance, online storage services such as Amazon S$3$, or IBM COS, only provide
server-side  encryption for protecting objects at rest~\cite{s3, cos}.  Similar words can be
said for  the access control of individual objects\footnote{In general terms, cloud object stores enforce access at the container level
rather than at the object level.},  which is currently either realized via simple object ACLs (Access Control Lists) as in S$3$~\cite{s3},
or not possible at all as in OpenStack Swift~\cite{biswas15}. 

This poor interface is insufficient for many applications.  For instance, 
it does not  enable in-place queries on encrypted data,  transparent and secure data sharing, 
and access control based on the content of an object. But also, it is 
very difficult to make it evolve to meet the changing needs of applications
and withstand  the  test  of  time. In practice,  most of these object storage systems leave no 
other choice to modifying the system internals to incorporate new security mechanisms
at the object level. This requires a deep knowledge of the system,  extreme care
when modifying  critical  software that  took years  of  code-hardening 
to trust,  and significant cost and time (see, for instance,~\cite{li19},  where it is described the ``daunting'' task 
of deploying new erasure coding solutions in object stores such as OpenStack Swift~\cite{swift} and Ceph~\cite{weil06}).

Rather than relying on object storage systems to  change,~we advocate in this paper
to ``work around'' the traditionally rigid  object storage APIs by embracing a  \emph{software-defined}
approach.  Similarly to software-defined networking (SDN),  we argue that
the separation of the ``control logic'' from the ``data protection  logic'' 
can give  the  needed flexibility  and ease of use to  enable users,   
programmers and sysadmins  to custom-fit access control
and object protection.  To give an example, pretend that  
upon certain conditions,  parts of an encrypted object need to be re-encrypted to share~it with the mobile
users of the application. Further, these conditions may depend on the contents of the object itself (e.g., on sensitive data such as sexual
orientation),  which must always remain confidential from the server. 
Simply put, what we pursue is  to offer users the ability to succinctly express this behavior at the object storage level,
and enforce~it by calling the corresponding  re-encryption modules. 

~Nevertheless, a software-defined  solution to enhance object storage data protection 
requires solving several issues at~two levels:

\begin{itemize}
\item  At the control plane,  by enabling the composition of per-object protection services. These compositions should be 
expressed concisely and in a manner agnostic to the data protection code and the remaining storage stack.
\item At the data plane, by making it truly programmable. To put it baldly, the data plane should not only allow to plug-in new protection
logic in the critical I/O path, but to run it safely. In addition, it should enable the re-usability of the protection capabilities,
so that users can compose new data protection controls.
\end{itemize}

In this research work, we present \egeon, a novel software-defined data protection framework for object storage. 
\egeon\hphantom{} exports a scripting API to define privacy policies for protecting objects. These policies enable data owners to declaratively 
set complex data protection services through the composition of user-defined transformations run in a serverless fashion.  
These functions represent the elementary processing units in \egeon, 
and may be re-used and linked together to implement complex privacy policies. The major feature of these transformations is that
they  are executed ``inline'' by \egeon~upon a standard~\texttt{Get} request. 
In this way, users can trivially display multiple views from the same object, and
modify these views by updating the policies. If some functionality to 
implement a view is missing, \egeon~provides a
simple API to deploy new transformations and customize the access to data objects. 

In this sense, one of the primary contributions of \egeon~is the ability to provide privacy-compliant transformed views of 
the underlying data on the fly. Perception of privacy can vary broadly
across applications. As an example, a dataset created by a hospital may include personally identifiable information (PII) that is not needed when it is
processed by a data analytics engine. Nonetheless, if the same dataset is accessed by medical personnel, a richer view of  it should
be given. Thus, a practical system needs to support a range of privacy preferences. 

By adopting a software-defined storage architecture, \egeon\hphantom{}
allows users to express their privacy preferences as policies~in the control plane and produce views
conforming to the policies by running transformations in the data plane. In this work, 
we focus on (cryptographic) transformations that process data as streams, that is, as
data is being retrieved from object storage nodes. Consequently, the first bytes of transformed views are
received as soon as possible, which permits \egeon~to scale to arbitrary object sizes without important
penalty on end-to-end latency.

The  \egeon~prototype we  present in this research has  been  implemented atop
OpenStack Swift~\cite{swift}. We took Swift~because it is open source and a  production quality
system.  Its sizable developer community ensures that  our new properties are  built on
code that is robust and that will be soon  evaluated. It must be noticed
that the design concepts underpinning \egeon~are generic and
could be easily ported to other storage substrates. 
For example, object 
classes allow to extend Ceph by loading custom code into Object Storage Daemons (OSDs),
which can be run from within a \texttt{librados} application~\cite{librados}. Thus, with some effort, it would be possible to leverage object classes to
implement transformed views of data.

~~Our performance evaluation of \egeon~shows promise in developing data
protection services with low overhead directly into the object storage. For instance, the overhead~of 
a \texttt{NOOP} policy, where a storage function simpy echoes the input data, is of around $9$ ms. Also,
depending on the amount~of protected data filtered out in the transformed views,~end-to-end
latency can be even lower with \egeon~due to the savings in network communication (up to $72.1$x for 
a $4$G mobile use case).

\begin{figure}
\includegraphics[width=0.5\textwidth]{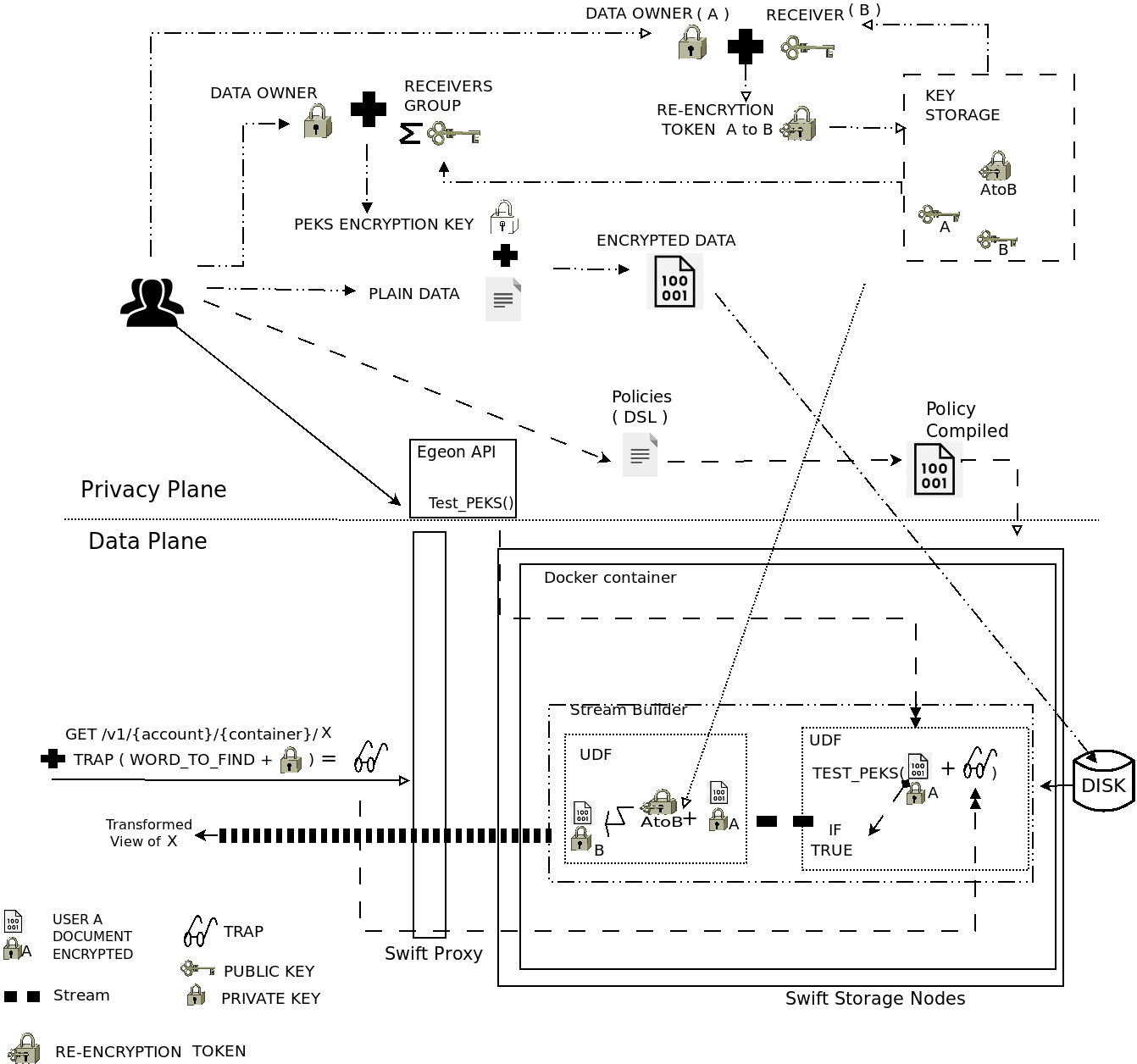} 
\caption{\egeon's architecture.}
\label{fig:architecture}
\vspace{-20pt}
\end{figure}

\section{Design}

\egeon~is a software-defined data protection framework that augments
object storage with composable security services to enforce users' privacy
preferences over protected data. These services are built up as
pipelines of serverless functions. Fig.~\ref{fig:architecture} illustrates an overview of \egeon's
architecture. Our objective is to enable authorized users or applications to access protected
data without violating the privacy policies of data owners.~We 
have designed \egeon~to make it easy the leverage of state-of-the-art privacy 
solutions  (such as content-level access control, homomorphic encryption, 
encrypted keyword search, ...) while preserving the normal data flow in the consuming applications. 
Concretely, we achieve this by introducing a logical separation between the \emph{privacy plane}, where
data owners set their privacy preferences, and the \emph{data plane}, where the creation of privacy-compliant transformed views
happens. This separation allows for heterogeneous policies atop the same
data without having to modify the system internals to enforce advanced policies at
the object level. 

\egeon's architecture consists of the following components:

\medskip
\noindent\textbf{Privacy Plane}.~~The privacy plane in \egeon~corresponds to the control plane of a software-defined 
architecture~\cite{crystal, FGCScrystal}, but specialized for data protection.  In practice, this means that \egeon~provides 
its own script language to assist data owner~in composing data protection services from elementary serverless functions
in the \emph{data plane}. The textual, JSON-based language supports conditions and compositions to build up inline privacy transformers (e.g.,
see Listing~\ref{lst:1}). Moreover, \egeon~offers an API to allow data owners to manage the life-cycle of their~data protection policies.
For performance reasons, once uploaded, the policies are automatically compiled into Java bytecode and stored in the
\emph{Metadata Service}. For fast access, this service has been built on top of Redis~\cite{redis}, an in-memory, low-latency key value store.  

\medskip
\noindent\textbf{Data Plane}. In \egeon, the data plane has a critical role. The data plane is responsible for
generating the privacy-compliant data views. And hence, it must be extensible to accommodate new 
functionality that enables privacy transformations on data. Particularly, in this realization of \egeon,
we have focused~on inline privacy transformations as data is retrieved from storage nodes HDDs. As mandated by
the policies in the privacy plane, privacy transformations are constructed from pipelines of user-defined functions
executed as serverless functions by \egeon. Namely, a user integrating a new transformation only needs to
contribute the logic. Resource allocation and execution of the chain
of transformations is automatically handled by \egeon, bringing a true serverless experience to users.

To minimize execution overhead, since many cryptographic operations are CPU-intensive,  \egeon~abides by
the principles of reactive programming and runs a transformation only when is strictly needed, instead of 
continuously on the data streams. More concretely, \egeon~extends the observer pattern~\cite{observer} and execute a certain
transformation in the pipeline when an event occurs~\cite{event}. Consequently, \egeon~better utilizes the available resources in 
the storage servers by balancing the load across the chain of transformations. To better understand this, pretend
that a user wants to compute the average salary of employees in a department $\mathtt{X}$. Now suppose that all the employee
records have been saved in a single JSON document with all the salary values homomorphically encrypted. Thanks to
\egeon~reactive core, the transformation to average the salary will only be run 
when the event ``$\mathtt{employee\,of\,department\,X}$'' comes through the data stream, thereby saving CPU resources.

\subsection{Threat Model}

Specifically, \egeon~enforces user's privacy preferences via function composition.
That is, users are ensured that their data is transformed as it goes through
a pipeline of transformation functions before it is released to applications. 
In the meantime, the original data remains end-to-end encrypted.

We assume an \emph{honest-but-curious}~\cite{cryptdb} storage servers, i.e.,
the server performs the computations correctly but will analyze all 
observed data to learn as much information as possible. We also presume the existence of an identity service (IDS) such as OpenStack Keystone for user authentication. 
We leverage~this service for authentication of the storage functions. An IDS is~a standard requirement in multi-user systems
and can even be a trustworthy external entity.

~Consumers of shared data are semi-trusted, in that they~do not
collude with the servers to leak the data or keys. This~is~a reasonable assumption for groups of data consumers that
are acquainted with each other. Further,
 \egeon~assumes that~the applications behave correctly and do not hand out user keys
to malicious parties. Finally, we assume state-of-the-art security
mechanisms to be in place for user devices, and that all parties
communicate over secure channels.

In this setting, \egeon\hphantom{} enforces \emph{data confidentiality}, making sure that 
the adversary learns nothing about the data streams, except what can be learned
from the transformed views. 

\smallskip
\noindent\textbf{Robustness.}~While \egeon~is able to handle various types~of failures 
in practice, provable robustness against misconfigured, or even malicious
privacy policies, and data producers is out of scope for this
research. A malicious user sending corrupted tokens cannot
compromise privacy but could alter the output~of a 
transformed view.

\subsection{Privacy Plane}
\label{sec:privacy}

~In the privacy plane, \egeon~provides the capabilities for data owner to set their privacy preferences ---i.e., user-centric privacy---, 
and what transformations will be required to apply to enforce a privacy policy. These transformations are specified by their unique
name, and their existence is verified when the privacy policy is to be compiled. A privacy policy applies to~a single object. In this
paper, we do not consider the question of how to set privacy policies for group of objects and how~they should look like. This
question has been left for future work.

In \egeon, targeted data objects in the privacy policies are specified by its full resource path. Following OpenStack Swift specs~\cite{swift}, the access path to an 
object is structured into three parts: ${\small \mathtt{/account/container/object}}$. As an example, for the ${\small \mathtt{rose.jpg}}$ object in the ${\small \mathtt{images}}$ 
container in the $1234$ account, the resource path is: ${\small\mathtt{/1234/images/rose.jpg}}$. 

Data owners can translate their preferences over an object to a set of transformations by mapping them in a JSON-based schema language. 
In addition to some meta-information (e.g., policy identifier), this schema permits data owners to formalize conditions at the policy level
using a rich set of operators such as ${\small ``\mathtt{StringLike}"}$, $\small ``\mathtt{NumericLessThan}"$, etc. Importantly,~the language
enables data owners to build date expressions using operators like $\small{``\mathtt{DateNotEquals}"}$, which makes it possible~to
express temporal restrictions. For instance, pretend that a data owner wishes to prevent that a document can be accessed~on weekends. 
She could indicate this through the date expression: ${\footnotesize \mathtt{``DateNotEquals": \left\{``Day": [``Sat",``Sun"]\right\}}}$.

More interestingly, this schema allows composing complex data protection transformations from elementary UDFs. This can be achieved
by adding each individual UDF as a step~in~the transformation pipeline defined in the JSON object ${\small ``\mathtt{Action}"}$. 
This object contains two name-value pairs: ${\small ``\mathtt{StartAt}"}$, which indicates the first transformation in the pipeline, and
${\small ``\mathtt{Steps}"}$, which is another JSON object that specifies the transformation UDFs along with their input parameters. 
Since transformations run only when the corresponding events come through the data stream, this schema allows data owners to
specify the observed events for each transformation to execute. To do so, there exists a field named  ${\small ``\mathtt{EventType}"}$
to signal the event to be observed by a particular transformation UDF. A typical ${\small ``\mathtt{EventType}"}$ block looks like
this:
\begin{lstlisting}[language=jsonnonumber]
{
  "Type": "<type_of_event>",
  "Input": [<parameter_block>, ...],
}
\end{lstlisting}
where the ${\small ``\mathtt{Type}"}$ field specifies the type of an event (e.g., an XPath~\cite{xpath} event) and 
the array ${\small ``\mathtt{Input}"}$ lists the parameters that are required for this type of event. For instance, if
a data owner wanted to apply a transformation UDF over all $\small \mathtt{salary}$ elements of an XML document, 
she could do so by setting an XPath event as follows:
\begin{lstlisting}[language=jsonnonumber]
{
  "Type": "XPathEvent",
  "Input": [{"Predicate": "//salary"}],
}
\end{lstlisting}
Similarly, a transformation UDF block is defined as follows:
\begin{lstlisting}[language=jsonnonumber]
{
  "Id": "<identifier_of_UDF>",
  "EventType": "<event_block>",
  "Input": [<parameter_block>, ...],
  "Next": (<identifier_of_UDF>|"End")
}
\end{lstlisting}

The ${\small ``\mathtt{Id}"}$ field is a string which uniquely identifies the UDF, 
while the ${\small ``\mathtt{Input}"}$ field permits to specify the parameters
for the transformation UDF (e.g., the targeted security level of a cryptosystem). Finally, the
${\small ``\mathtt{Next}"}$ field indicates the next step to follow in the pipeline,
or ${\small ``\mathtt{End}"}$ to indicate the end of the chain of transformations.

An example of a real policy can be found in Listing~1. This policy provides
transformed views over a JSON file containing employee records of the format:
\begin{lstlisting}[language=jsonnonumber]
{
  "employee": {
	"name": "Alice",
	"identification": {
	  "SSN": "32456677"
	},
	"salary": 50000
  }
  ...
}
\end{lstlisting}
As a first transformation in the chain, this policy uses
content-level access control~\cite{biswas15} (CLAC). Very succinctly, CLAC works as follows. It assigns targeted JSON elements
an object label (${\small \mathtt{olabel}}$) and each user~a user-label ($\mathtt{ulabel}$). Then, it allows to define rules in the form~of $(\mathtt{ulabel}, \mathtt{olabel})$,
which means that the JSON elements labeled with any of the $\mathtt{olabel}$s are allowed to be read 
by the users labeled with the corresponding $\mathtt{ulabel}$s. To indicate what JSON items to protect,
CLAC uses JSONPath in this case.

 In this policy,  we assume two types of users:  treasurers with user label $``\mathtt{treasurer}"$
and regular users with label ${\small ``\mathtt{user}"}$. We protect the salaries with the object
label $``\mathtt{sensitive}"$ and specify the single rule $(``\mathtt{treasurer}", ``\mathtt{sensitive}")$,
which means that only treasurers will have access to the salaries. The field
$``\mathtt{salary}"$ is identified using the JSONPath expression:
$`\mathtt{\$.employee.salary}$' (Listing~1, line $14$).

The second transformation applies homomorphic encryption on the field 
$``\mathtt{salary}"$ to prevent the servers from learning the 
employee salaries~\cite{pilatus}, while computing the average salary of employees (Listing~1, lines $20$-$30$).

As a final transformation in the chain, the policy uses proxy re-encryption~\cite{pilatus} to convert the
homomorphically  encrypted average salary to a ciphertext under the receiver's key. Thus, the
transformed view can be decrypted by the receiver,  without the data owner having to share her
private key nor performing any encryption for the receiver on her personal device.

To wrap up, this policy will generate two transformed views of the same data.
For regular users, it will only be executed the first step (lines $10$-$19$): the CLAC transformation. 
Due to lack of permissions, the CLAC module will eliminate the encrypted $``\mathtt{salary}"$ field, 
and output a transformed view with the rest~of information. For a treasurer, it will output the same
view as~any regular user (without individual salaries), but enriched with the average salary encrypted under 
her public key thanks to proxy re-encryption (Listing~1, lines $31$-$37$).

We want to note that the different transformations UDF only execute when the corresponding JSONPath
events come along.  To wit, proxy re-encryption will only be run one time, after~the JSON field named $``\mathtt{average\_salary}"$  
is added at the end of the response by the second transformation in the chain.

\begin{lstlisting}[caption={A sample policy to process employee records.}\label{lst:1},language=json,firstnumber=1,captionpos=b]
{ 
 "Id": "employee.policy",
 "Object": "v1/{account}/{container}/employees.json",
 "Condition": { 
    "DateNotEquals": { "Day": ["Sat","Sun"] } 
 },
 "Action": {
  "StartAt": "Step1",
  "Steps": {
    "Step1": {
     "Id": "CLAC",
     "EventType": {
        "Type":  "JSONPathMarkerEvent", 
        "Input": [{"Predicate":"$.employee.salary", "olabel": "sensitive" }]  
     },
     "Input": [{"ulabel": "treasurer", 
                "olabel": "sensitive" }],
     "Next": "Step2" 
    },
    "Step2": {
     "Id": "SUM",
     "EventType": {
        "Type":  "JSONPathEvent", 
        "Input": [{"Predicate":"$.employee.salary"}]  
     },
     "Input": [{"average": true},
	       {"keyOwner": "meta://Alice/keys/hom",
	       {"output":"$.average_salary"}],
     "Next": "Step3"
    },
    "Step3": {
     "Id": "PRE",
     "EventType": {
        "Type":  "JSONPathEvent", 
        "Input": [{"Predicate":"$.average_salary"}]  
     },
     "Next": "End"
}}}
\end{lstlisting}

As a final word, this
example clearly displays how \egeon\hphantom{} is  capable of performing real-time 
privacy transformations~atop the same data object for  a variety of application scenarios,~thus
enhancing the rigid interface of object storage systems.

\subsection{Data Plane}
\label{sec:data}

~~The focus of \egeon~is on inline privacy transformations, where 
data streams are ``observables'' and user-defined privacy transformations 
are ``observers'' subscribed to the data streams.
As as soon as an event is observed, it will be delivered to the subscribed observers.
If there are no events on the data stream then the original data stream is pushed
back to the user. In~this sense, a privacy transformation is nothing but a function taking an
observable as input and returning another observable as its output. This design has
the advantage that transformations can be chained together to generate complex
data views compliant with the policies in the privacy plane.

\smallskip
\noindent\textbf{Runtime.} Currently, \egeon's runtime is Java-based. Thus,~the chain of
transformations is run within a Java Virtual Machine (JVM) wrapped
within a Docker container to guarantee a high level of isolation between
two different data transformations. 

\smallskip
\noindent\textbf{Policy Enforcement.}
~~Upon a new \texttt{Get} request,
\egeon~starts up a thread inside the JVM to perform three
tasks. We refer~to this thread as the ``master thread''. The three tasks in order
of execution are:
\begin{enumerate}
\item \emph{Policy loading}, where the master thread loads the policy into memory and evaluates its conditions clauses.
\item \emph{Observable setting}, where the master thread opens a~data stream to the target object, namely, the observable, if
the policy conditions are fulfilled. To this aim, it creates an instance of the appropriate subclass of the abstract class
$\mathtt{StreamBuilder}$ that the \egeon's engine leverages to start parsing the object.
  Subclasses are required since~the specific 
logic to parse the data stream and generate the events depends on the type of file.  
Specifically, \egeon\vphantom{} picks up 
the proper $\mathtt{StreamBuilder}$ subclass based on the object extension
(e.g., $``\mathtt{.json}"$ for~JSON documents) using \emph{factory} methods.
\item \emph{Transformations setting}, where the master thread makes an instance of each
transformation UDF in the pipeline and chains them together. The $\mathtt{StreamBuilder}$
subclass generates events as the data is being parsed, and notifies the subscribed
transformation UDFs in the same order as dictated in the privacy policy. To wit,
in Listing~\ref{lst:1}, upon the JSONPath event $`\mathtt{\$.employee.salary}$',
\egeon~will execute first the CLAC transformation, followed by the SUM transformation.
\end{enumerate}
In \egeon, we assume that an observable can only handle~one event at a time. We adopted
this design to minimize compute resources at the storage layer, so that both
the data stream and all its transformations operate in the same thread, in our case, the master
thread. Nevertheless, this can be easily changed by switching to a different thread and integrating some 
additional logic for coordinating the threads.

Since each transformation UDF acts as an observer, another responsibility of
the master thread is to subscribe each privacy transformation to the events
specified in the policy. To this aim, it invokes the method~${\tiny \mathtt{install(Event\,event, UDF\,observer)}}$ in
 the ${\small \mathtt{StreamBuilder}}$ class, where the parameter ${\small \mathtt{observer}}$ is the
transformation UDF bound to the ${\small \mathtt{event}}$.

\smallskip
\noindent\textbf{Extensibility.} At the time of this writing, \egeon~implements three types of data sources:  XML, JSON and CSV 
documents, a number of events including XPath and JsonPath expressions, CSV field and records, etc., and multiple transformation
UDFs (see \S\ref{sec:trans} for further details). However, \egeon~is extensible, and new events, observables and observers can be incorporated by
extending the abstract classes ${\small \mathtt{Event}}$, ${\small \mathtt{StreamBuilder}}$ and 
${\small \mathtt{UDF}}$, respectively. 

Due to space constraints, we only show here an example of a transformation ${\small \mathtt{UDF}}$ to
perform summations on ciphertexts~\cite{pilatus} to see how easy
it is to code a transformation UDF (Listing~2).

\begin{lstlisting}[caption={A transformation UDF to perform summations on ciphertexts.},language=Java, captionpos=b,	numbers=left, 
		numberstyle=\footnotesize]
package com.urv.egeon.function;

import com.urv.egeon.runtime.api.crypto.Homomorphic;
import com.urv.egeon.runtime.api.parser.UDF.ContextUDF;
import com.urv.egeon.runtime.api.parser.UDF.UDF;
import com.urv.egeon.runtime.api.parser.event.Event;

public class Sum extends UDF {
    private Homomorphic accum = new Homomorphic();
       		
    @Override
    public Event update(Event e, ContextUDF ctx) {
       try {
          this.accum.add((String) e.getValue());
       } catch (Exception ex) { 
          // first execution of the UDF
          this.accum.setKeys((String) ctx.getParameter("keyOwner"));
          this.accum.setCipher(this.accum.fromSerial((String) e.getValue()));
       } finally {
          return e;
       }
    }
    @Override
    public Object complete(ContextUDF ctx) { ... }
}
\end{lstlisting}

As shown in Listing~2, a UDF has two methods: the method $\mathtt{update}$, 
which is invoked every time a new subscribed event is emitted, and the method $\mathtt{complete}$, 
which is called when~the data stream is finalized (empty). The $\mathtt{update}$ method has two arguments
of type $\mathtt{Event}$ and $\mathtt{ContextUDF}$. The first argument encapsulates the details of the emitted event.
For instance, for the homomorphic summation UDF of Listing~2, this may mean a JSONPath event, alongside
the value of the selected field by the JSONPath expression (e.g., an encrypted salary value), and accessible via
the method $\mathtt{getValue}$ (Listing~2, line $14$). The $\mathtt{ContextUDF}$ encapsulates the access to the
request metadata, such as HTTP headers and cryptographic keys and tokens sent out by the client, 
the object's metadata, and specific parameters required for the UDF to work, such as the data owner's public key to
operate on the encrypted values (Listing~2, line $17$). All this information is automatically made available by
\egeon~to the transformation UDF. This includes the input parameters~set in the privacy policy (e.g., Listing~1, lines $26$-$28$).

~~It is worth to mention here that the input arguments whose values have~the format of
$``\mathtt{meta://key}"$ are automatically downloaded from the Metadata
Service using the key $``\mathtt{key}"$. In this way, a data owner can change the input parameters (e.g., her
homomorphic public key) without having to re-compile the policy. An example of this can be found in 
Listing~1, line $27$,  for the argument $``\mathtt{keyOwner}"$, whose value is retrieved from the Metadata
Service behind the scenes, and made accessible to the summation UDF via the context
(Listing~2, line $17$).

~~The purpose of the method $\mathtt{complete}$ is to provide a hook for 
developers to perform final computations and append their result at the 
end of the data stream. For instance, this could be useful to compute
the average over encrypted data and append the result as new item
at the end of a JSON document.

\subsection{Data Transformation UDFs}
\label{sec:trans}

\noindent\egeon~comes up with a library of reusable functions~(UDFs) for
inline data protection. While some of these functions apply cryptographic
transformations, others act on raw data, e.g., by filtering out protected 
parts of a JSON object to non-privileged users~\cite{biswas15}. All  functions are composable
to provide complex transformed data views. These
are the following:

\smallskip
\noindent\textbf{Homomorphic encryption (HOM).}~~HOM
is a cryptosystem (typically,  IND-CPA secure) that allows the server to perform
computations directly on encrypted data, the final result being 
decrypted by the user devices.   For general operations, HOM is prohibitively slow.~~However, it is
efficient for summation.~To support summation, along with proxy re-encryption (PRE), 
we adopted the homomorphic cryptosystem of~\cite{pilatus}. We chose this scheme,
because it allows secure data sharing and is tailored~to mobile platforms
and constrained IoT devices. Concretely,~we implemented two UDFs:
\begin{itemize}
\item Summation (\texttt{SUM}): This UDF supports the summation of ciphertexts, such
that the result is equal to the addition~of the plaintext values:$\,\mathtt{Enc}(m_1)\cdot\mathtt{Enc}(m_2)=\mathtt{Enc}(m_1 + m_2)$. 
\item Re-Encryption (\texttt{PRE}). Succinctly, \texttt{PRE} allows the storage
servers  to convert ciphertexts under the data owner's key to ciphertexts under authorized users keys without
leaking the plaintext. Therefore, the data owner can securely share data with other users, i.e., without sharing her
private key nor performing any encryption  for them on her personal device. 
To do so, the data owner $d$ solely needs
to issue a re-encryption token for a user $u$ based on his public key $pk_u$ as the $\mathtt{Token}_{d\rightarrow u}$.
With this token, the \texttt{PRE} UDF~can automatically re-encrypt data on behalf of the data owner without her
intervention.  Further, the re-encryption tokens are unidirectional and non-transitive.
\end{itemize}
For both UDFs, we assume integers of $\leq 32$ bits with $128$-bits of security. The implementation
makes use of the optimal Ate pairing~\cite{ate} over Barreto-Naehrigopera elliptic curve~\cite{bn}, and
also applies the Chine Remainder Theorem (CRT) to optimize
decryption~\cite{pilatus}.  For all this cryptographic processing,~we~use the RELIC toolkit~\cite{relic}.

\smallskip
\noindent\textbf{Keyword search (SEARCH).} To allow keyword searches (as the  SQL``\texttt{ILIKE}'' keyword), 
we make use of a cryptographic scheme for keyword searches on encrypted text~\cite{peks}. As above, we have
chosen this scheme because it is multi-user. To put~it baldly, before storing an object, the data owner first
selects~the users with whom she wants to share her data and then encrypts it with their public keys. To
search for keywords in the shared object, a user makes a trapdoor $\mathtt{Trap}_W$ for the keyword~set~$W$
using his private key, and then sends it to the server. The server runs the \texttt{SEARCH} UDF, which takes the
public key of the user, the trapdoor  $\mathtt{Trap}_W$ and the encrypted text, and returns ``yes'' if contains
$W$ or ``no'' otherwise. As expected, this scheme~is proved secure against chosen keyword attacks~(IND-CKA).

~One major advantage of this scheme is its short ciphertext size. Concretely, for $n$ users,  it requires 
$(n + \ell + 1) \cdot L1$ bits, where $\ell$ is the number of keywords and $L1$ is the bit length~of 
the underlying finite field $\mathbb{F}_q$, which is much smaller than the deployment of $n$ separate 
instances of a public key searchable encryption scheme, one for each user. For the implementation of this
scheme, we use the ``SS$512$'' elliptic curve, a symmetric curve with a $512$-bit base field, which provides
a security level of $80$-bits, from the jPBC library~\cite{JPBC}.

\smallskip
\noindent\textbf{Content-level access control (CLAC).} As an example of a non-cryptographic function,
we decided to implement content-level access control~\cite{biswas15}. The central idea of \texttt{CLAC}
is to enforce access control at the content level to restrict who reads which part 
of a document. To give a concrete example, consider that a hospital stores its patient records
as big JSON object. These records should be accessed
differently by different personnel. For example,  a ``doctor'' could
see health information from her patients, while a ``receptionist'' should only
view basic profile information about the patients.  With \texttt{CLAC}, a data owner
can define content-level policies to censor access to parts of a data object.  Remember that
in object storage systems such as Swift or Amazon S$3$, once an object is made accessible to someone, 
she retrieves the full content of the object. So, there is no way to hide out sensitive information to that user.

Our implementation of \texttt{CLAC} overcomes this limitation. As introduced in \S\ref{sec:privacy},
\texttt{CLAC} borrows the LaBAC model  (Label Based Access Control Model)~\cite{biswas15}. Either
be an XML element, a JSON element, or a CSV column, an object label (${\small \mathtt{olabel}}$)~is 
assigned on the targeted item. Similarly, each authorized user is given a user label  ($\mathtt{ulabel}$). Then,
the LaBAC model works by specifying tuples of rules in the form~of $(\mathtt{ulabel}, \mathtt{olabel})$,
which tell that only the users labeled with $\mathtt{ulabel}$ can access the items labeled with that $\mathtt{olabel}$. 
For instance,  if only users with the user label $``\mathtt{manager}"$ were authorized to access items labeled with $``\mathtt{restricted}"$,
a data owner should have to set the rule $(``\mathtt{manager}", ``\mathtt{restricted}")$ to effectively formalize access control. 
As in the LaBAC model, our implementation admits hierarchies of both $\mathtt{ulabel}$s and $\mathtt{olabel}$s to
rank users and objects.

~One interesting side effect of our reactive data plane is that our \texttt{CLAC} implementation is generic,
and not tied to a specific file type. What changes is the mechanism to identify the items, which depends
on the object type. That is, for XML, an XPath expression should be used to indicate that a certain
element has $``\mathtt{restricted}"$ access, while for a JSON document, a JSON predicate should be used in its stead.
We have abstracted this coupling by the definition of specific marker events as shown in Listing~1, lines $22$-$25$.

\section{Implementation}
\label{sec:impl}

We have constructed \egeon~by extending Zion~\cite{zion}, a 
data-driven serverless computing middleware for object storage. In
particular, we have implemented \egeon~on top of OpenStack Swift~\cite{swift},
a highly-scalable object store. 

OpenStack Swift is split into several components. The main components
are the object and proxy servers. While the object servers are responsible for the storage and management of the objects,
the proxy servers expose the RESTful Swift API (e.g., $\mathtt{GET\,/v1/account/container/object}$ to get object content)
and stream objects to and from the clients upon request.

To be as non-intrusive as possible, the only modification we perform in the default Swift architecture is the
deployment of~a custom Swift middleware to intercept $\mathtt{GET}$, or read, requests~at the proxy servers~\cite{middleware}.
This middleware also provides a simple API to manage the life-cycle of privacy policies. Essentially, it communicates
with the Metadata Service to store and retrieve the privacy policies for protected data objects. Recall that the Metadata
Service leverages Redis~\cite{redis} to yield sub-millisecond access latency to metadata.
To optimize request matching even further, we have collocated the Redis instance with the proxy server. 

~~As Zion, \egeon~uses containers to sandbox the execution of the chain of privacy transformations. 
However, contrary to Zion, which is a general-purpose serverless platform,  \egeon\hphantom{} employs 
a single, optimized serverless function to produce the privacy-compliant views of the 
underlying data. This function deserializes the compiled privacy policy, loads it into memory, 
evaluates the conditions from the clauses, and if it ``applies'',~it runs the the pipeline of UDF transformations.
This design has two main benefits. On the one hand, \egeon~runtime starts up faster. On the other hand, 
UDF transformations enjoy of a great level of isolation. Simply put, they have neither direct network access nor access to the local Linux file
system, among other namespaced resources, which protects the whole system from malicious transformations.

Also, to enhance response time for policies that are accessed frequently, \egeon~runtime deploys a cache for policies
using the Google Guava caching library~\cite{guava} configured with a least-recently-used (LRU) eviction policy.

~~Resource allocation is managed  by Zion. \egeon~does not contribute any optimization at this level.  
When a read request comes along, our Swift middleware contacts the Zion service, which manages
the containers in the object servers, and starts up a new one if necessary. 

\section{Evaluation}

~In this section, we evaluate key aspects of \egeon's~design and our prototype implementation. 
First, we begin with a series of microbenchmarks to judge aspects such as system overhead,
throughput and the performance of the cryptographic operators in isolation. Finally, we
assess \egeon's flexibility to compose complex data views.

\smallskip
\noindent\textbf{System setup.} All the experiments have been conducted in a cluster of
$8$ machines: $2$ Dell PowerEdge R320 machines with $12$GB RAM, which operate as Swift Proxy servers,  
and $6$ Dell PowerEdge R320 machines with $8$GB RAM and $4$-core CPUs, which act as
object servers. The version of  OpenStack Swift~is Stein $5.2.0$.  All the machines run 
Ubuntu Server $20.04$LTS~and are interconnected through $1$GbE links. The client machine for the experiments is 
equipped with a Intel Core i5-4440 CPU with 4 cores and $8$GB RAM.

\smallskip
\noindent\textbf{Competing systems}. In some tests, we have compared \egeon\hphantom{} against plain Zion~\cite{zion} 
and Vertigo~\cite{vertigo}, all deployed in the same Swift cluster as above. Concretely, we have used  Zion~as a baseline
to assess the overhead added by \egeon's software-defined architecture to the original Zion design, while we have chosen Vertigo as an 
example of a general-purpose, software-defined object storage system. As \egeon, Vertigo allows users to create pipelines 
of storage functions, each implemented as an OpenStack Storlet~\cite{storlet}. Hence, it is a good representative to act as
a proving ground for the performance of \egeon~against a similar software-defined architecture. It must be noticed that
the control plane in Vertigo is programmatic, while in \egeon\hphantom{} is declarative through the use of JSON-based privacy policies.

 \subsection{Microbenchmarks}

We have run several microbenchmarks:

\medskip
\noindent\textbf{Cryptographic operations.} 
To better understand the sources of overhead incurred by \egeon, we 
examined the throughput of the individual transformation UDFs, since
different privacy policies may result in various transformation mixes.
For each type of cryptographic transformation, we measured the number
of operations per second that the \egeon~runtime can perform on a 
object server in the data plane, as well as the latency.~The meaning of each operation depends on 
the specific type of~the cryptographic transformation UDF. 
For \texttt{HOM}, this refers to~the
 summation of two encrypted $32$-bit integers. For \texttt{PRE}, it refers to the proxy re-encryption of
a single encrypted $32$-bit integer, while for \texttt{SEARCH}, it represents the
search of a keyword in~an encrypted document with the same keyword. The results of this experiment are given in 
Table~\ref{tab:OPsSecond}. As expected, we can observe that the latency of the cryptographic operations is 
in the order of a few milliseconds, which is acceptable for many use cases. Due to the added latency of the cryptographic
transformations,  a reactive data plane such as that available in \egeon, which~is driven exclusively 
 by the events appearing in the data streams, can be of great help to define privacy policies that minimize~the
number of cryptographic operations (for instance, by skipping unneeded data in the first steps of the
transformation chain).

\begin{figure*}
\centering
\subfloat[][TTFB for a $10$KB object.]{
\includegraphics[width=0.325\textwidth]{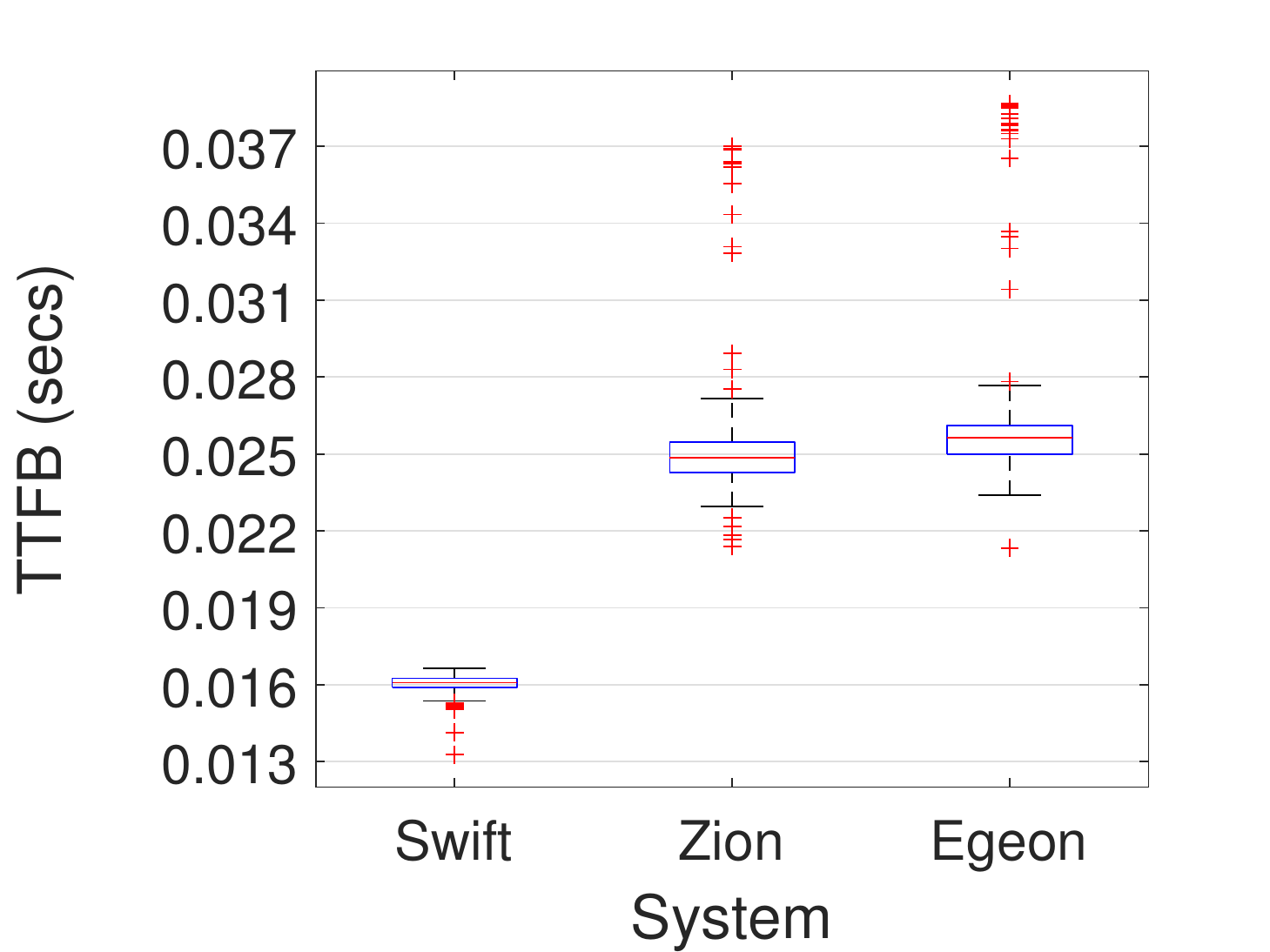}
}
\subfloat[][TTFB for a $100$KB object.]{
\includegraphics[width=0.325\textwidth]{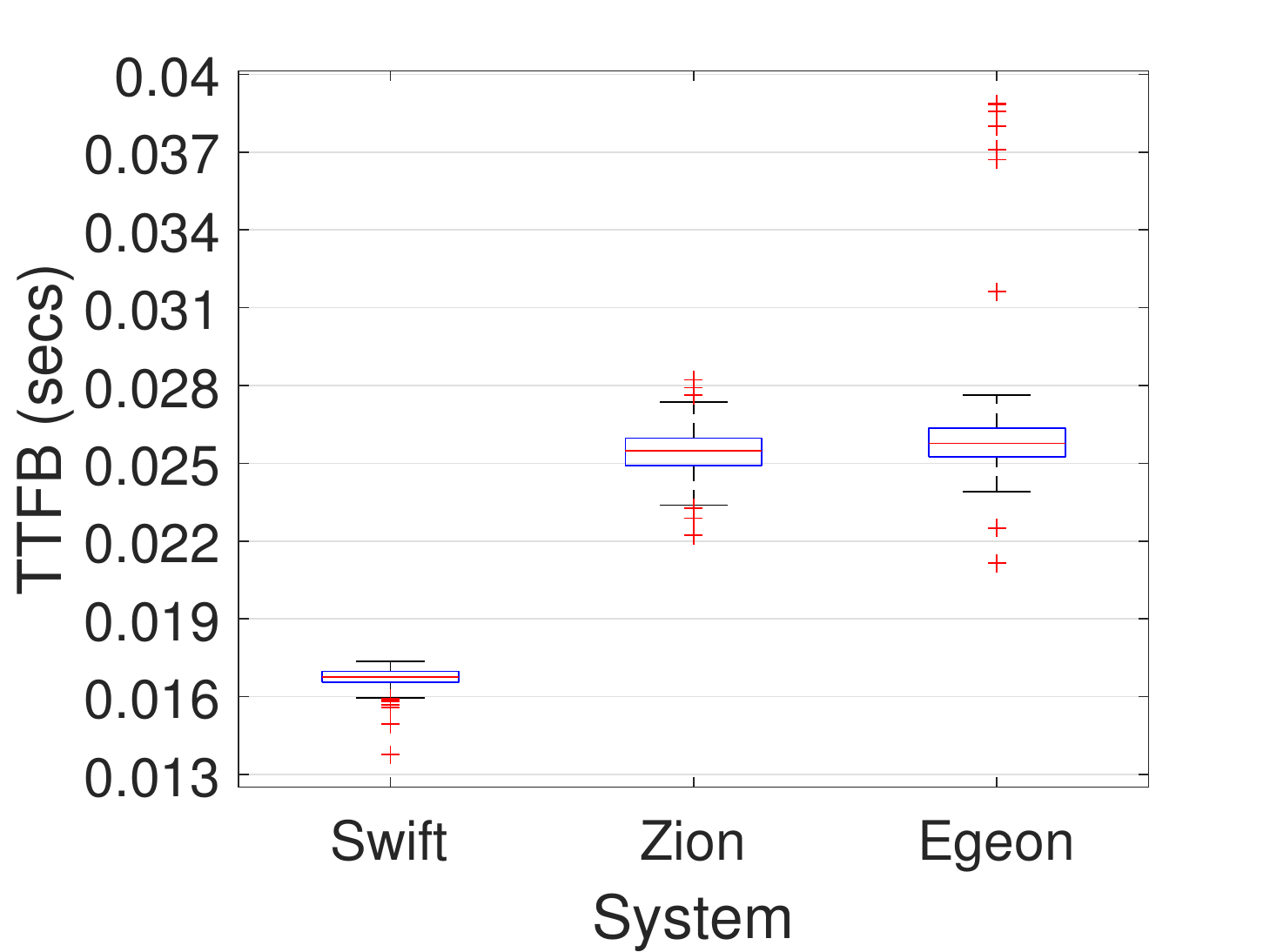}
}
\subfloat[][TTFB for a $1$MB object.]{
\includegraphics[width=0.325\textwidth]{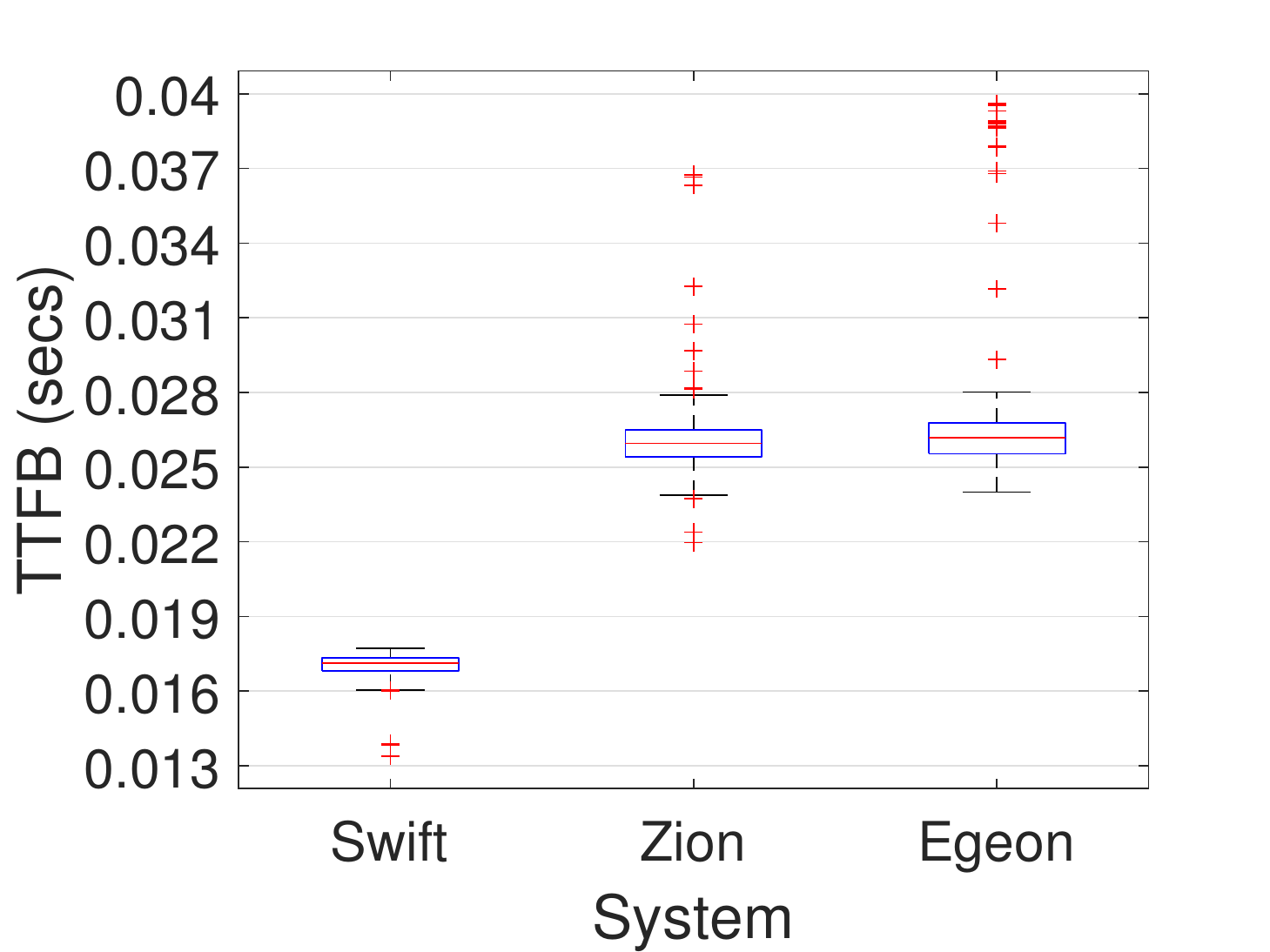}
}
\caption{Time to First Byte (TTFB) for different object sizes.}
\label{fig:2}
\vspace{-10pt}
\end{figure*}

\medskip
\noindent\textbf{Overhead.} To provide a full picture of the overheads incurred by \egeon, we
measured the latency introduced by \egeon~to the I/O path with respect to Zion and vanilla Swift.
To measure the overhead, we utilized the Time to First Byte (TTFB),~which captures how long the 
client needed to wait before receiving~its first byte of the response payload from the Swift servers. To
make measurements more precise, we colocated the client with one of the Swift proxy servers. As a client,
we used \texttt{pycurl} to generate the \texttt{Get} requests for three different data file sizes.
For each object size, we performed $1$K requests. For Zion,~each request caused the invocation of a \texttt{NOOP} function
that simply echoes the input stream to the output (see Listing~1 in \cite{zion}~for further details).  For \egeon, we 
set up a \texttt{NOOP} policy which includes a single \texttt{NOOP} UDF in the transformation chain. This UDF
does nothing: 

\smallskip
\begin{lstlisting}[caption={A no-operation (NOOP) UDF.},language=Java, captionpos=b,	numbers=left, 
		numberstyle=\footnotesize]
package com.urv.egeon.function;

import com.urv.egeon.runtime.api.parser.UDF.ContextUDF;
import com.urv.egeon.runtime.api.parser.UDF.UDF;
import com.urv.egeon.runtime.api.parser.event.Event;

public class Noop extends UDF {
    @Override
    public Event update(Event e, ContextUDF ctx) { return e; }
    @Override
    public Object destroy(ContextUDF ctx) { return null; }
}
\end{lstlisting}
The results are given in Fig.~\ref{fig:2}. As seen in this figure, Zion and \egeon~are on par, which
demonstrates that \egeon\vphantom{} software-defined architecture adds little overhead to Zion.
 With respect to vanilla Swift, both systems add around $9$ms of extra latency, which can be
considered very small. To wit, serverless function invocation in major cloud providers usually take
between $25$ to $320$ms in warm state~\cite{peeking}.

\begin{figure}
\centering
\subfloat[][Overhead of \egeon.]{
\includegraphics[width=0.245\textwidth]{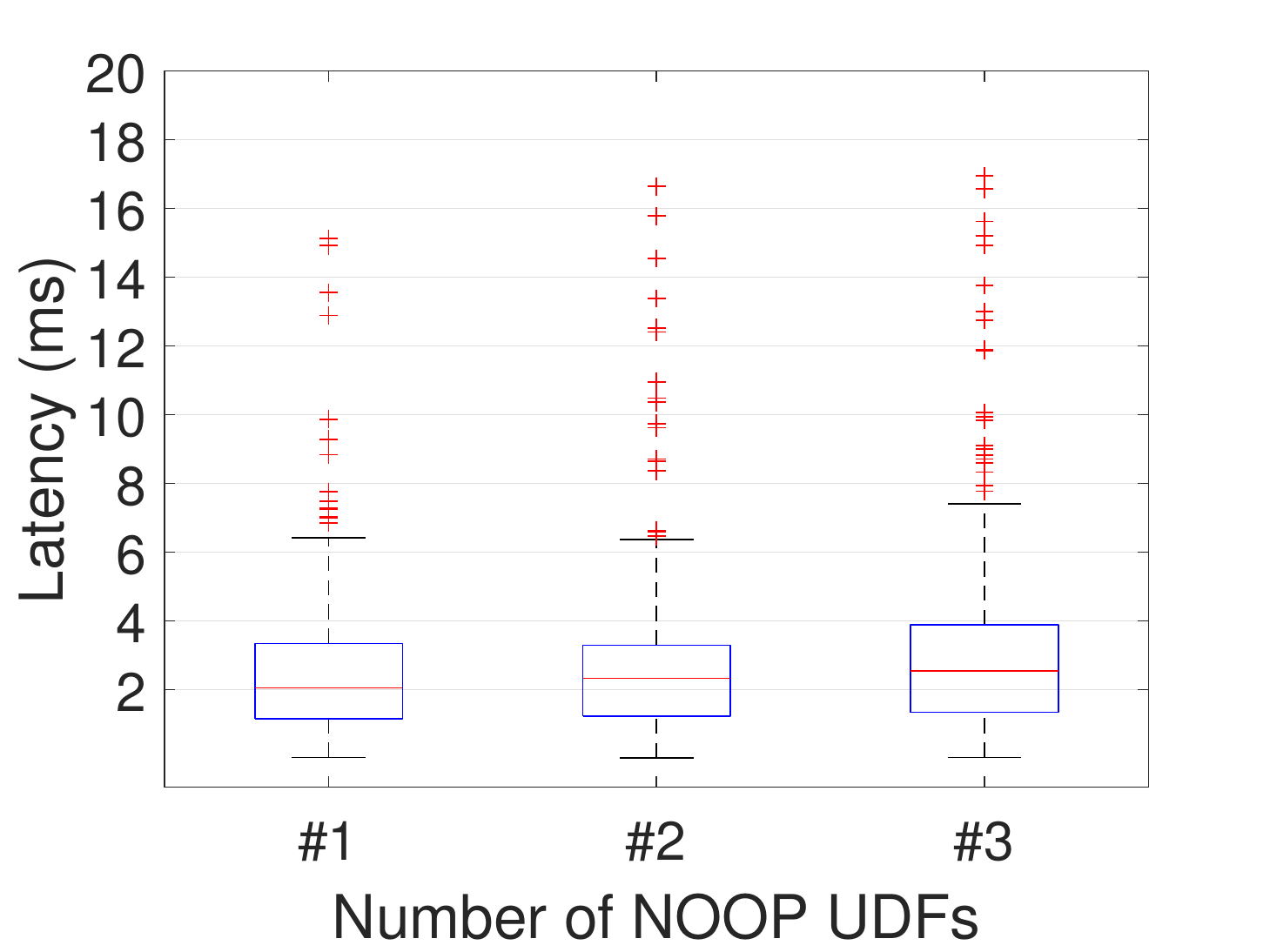}
	\label{fig:3a}
}
\subfloat[][Overhead of Vertigo.]{
\includegraphics[width=0.245\textwidth]{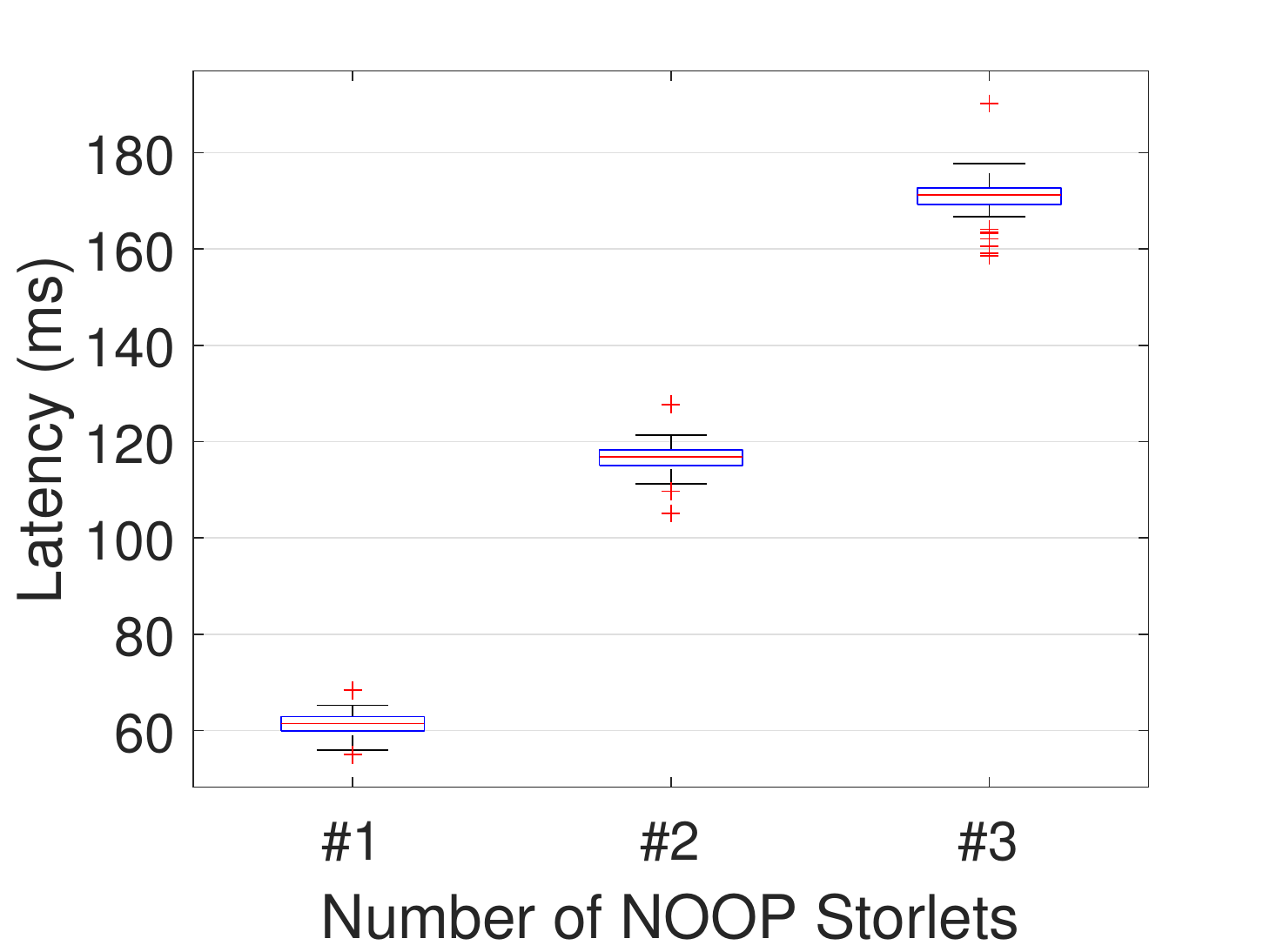}
\label{fig:3b}
}
\caption{Overhead of chain setup of \egeon~versus Vertigo.}
\vspace{-5pt}
\label{fig:3}
\end{figure}

Since Zion does not support the pipelining of functions, we compared \egeon~against Vertigo.
Recall that Vertigo enables users to chain several Storlets together, where each Storlet~can
implement some reusable storage function such as decryption, compression, etc. We repeated
the same experiment as above, but evaluating chains of increasing length. For \egeon, we
set up chains of \texttt{NOOP} UDFs, while for Vertigo, we did the same, but for pipelines of 
\texttt{NOOP} Storlets. The results are depicted~in Fig.~\ref{fig:3}. We can see that while the
overhead keeps constant in \egeon, Vertigo shows a linear increase in latency. The reason
for such a difference is the reactive core of \egeon,~which~does nothing if the
transformations are not subscribed to any event. Vertigo, however, wires each
Storlet with their neighbors in~the chain, which takes some time, albeit each of them just
copies the input to the output. This strongly reinforces the idea that for a software-defined
data protection system to be useful, it~is not a good idea to route the data streams through
a pipeline~of functions, but rather to act on them when it is strictly needed. Indeed, Vertigo's
overhead is more than one magnitude higher than \egeon's overhead as shown in Fig.~\ref{fig:3}.

\begin{table}[t!]
\caption{Performance of cryptographic primitives available in \egeon.}
\label{tab:OPsSecond}
 \begin{center}
 \begin{tabular}{|l|l|l|}
 \hline
 \textbf{Transformation UDF} &  \textbf{Throughput} (ops/sec)   & \textbf{Latency} (ms)\\  \hline\hline
Homomorphic Addition (\texttt{SUM})	&	$616$ & $1.62$ \\ 
Proxy Re-Encryption (\texttt{PRE})  & 137 & $7.29$ \\  
Keyword Search (\texttt{SEARCH}) & 166 & $6.02$ \\  \hline
\end{tabular}
\end{center}
\vspace{-20pt}
\end{table}

\medskip
\noindent\textbf{Throughput.} As a final microbenchmark, we quantified the impact of \egeon~on
the system throughput. As above, \egeon\vphantom{} was compared to Zion and vanilla Swift to 
give real sense of its performance. For this experiment, we utilized the \texttt{getput} benchmarking tool suite~\cite{getput} for Swift.
And in particular, the \texttt{gpsuite} to conduct parallel tests with multiple clients. More concretely, we run \texttt{gpsuite} in one of
the Swift proxy servers for $10$ seconds and for different object sizes. We considered a replication factor of $3$, and
instrumented \texttt{gpsuite} to stress $3$ out of the $6$ object servers in the data plane. As in the overhead test, a \texttt{NOOP}
function call per request was made for Zion and a \texttt{NOOP} policy plus \texttt{NOOP} UDF for \egeon. Table~\ref{tab:IOPS}
reports the maximum throughput in operations per second attained by each system. Similarly to what was observed for the overhead,
\egeon~and Zion perform in similar terms. More interestingly, as the object size increases, the gap between both \egeon~and Zion and Swift
grows. We investigated this issue and we found that this happens due to a higher CPU interference caused by the JVM used to run
the \egeon~logic and the \texttt{NOOP} code~in Zion, respectively.  

\begin{figure*}
\centering
\subfloat[][Throughput (ops/sec) for a $1$MB object.]{
\includegraphics[width=0.3\textwidth]{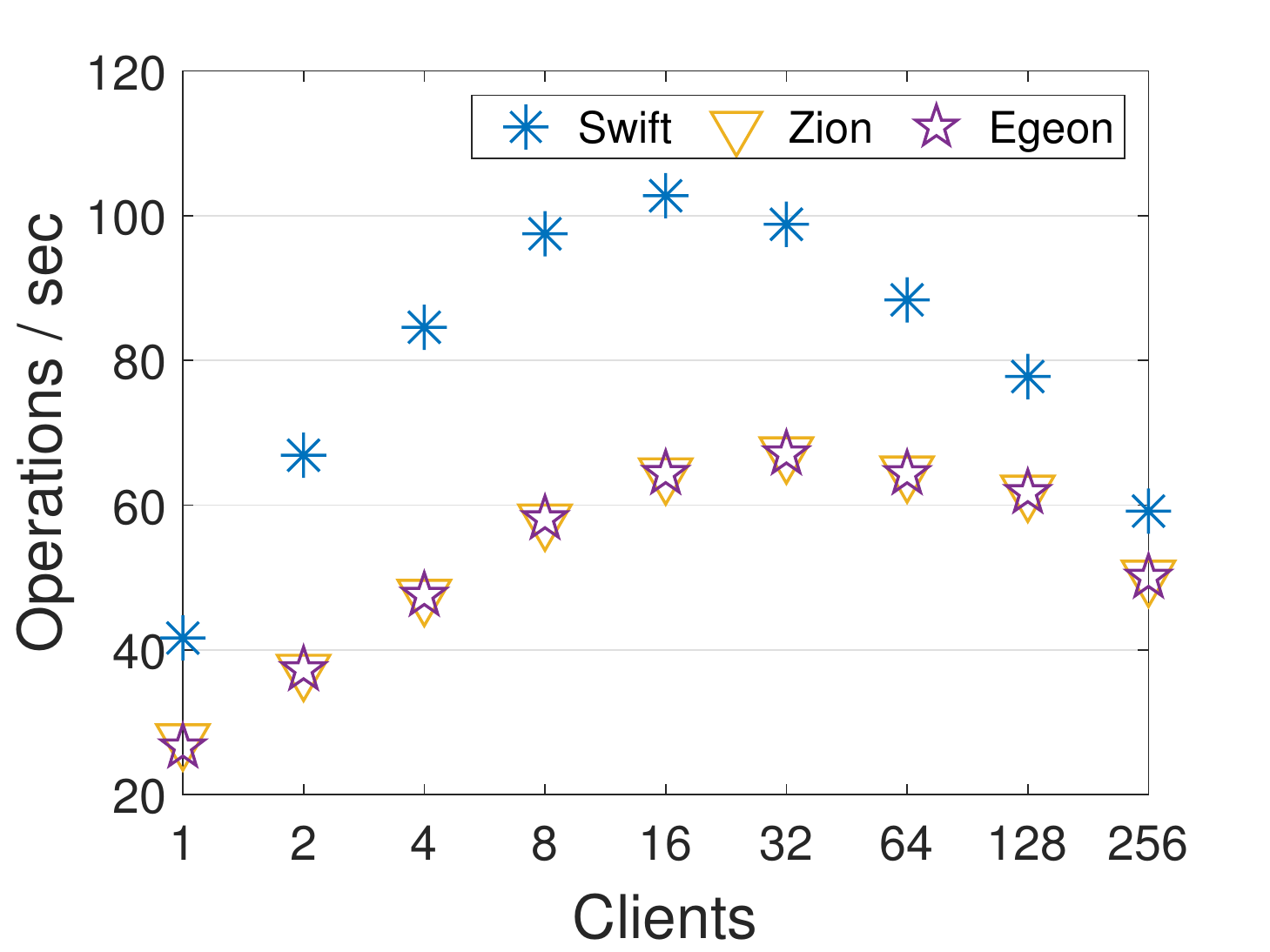}
	\label{fig:4a}
}\quad\quad\quad
\subfloat[][Slowdown of \egeon~over vanilla Swift.]{
\includegraphics[width=0.325\textwidth]{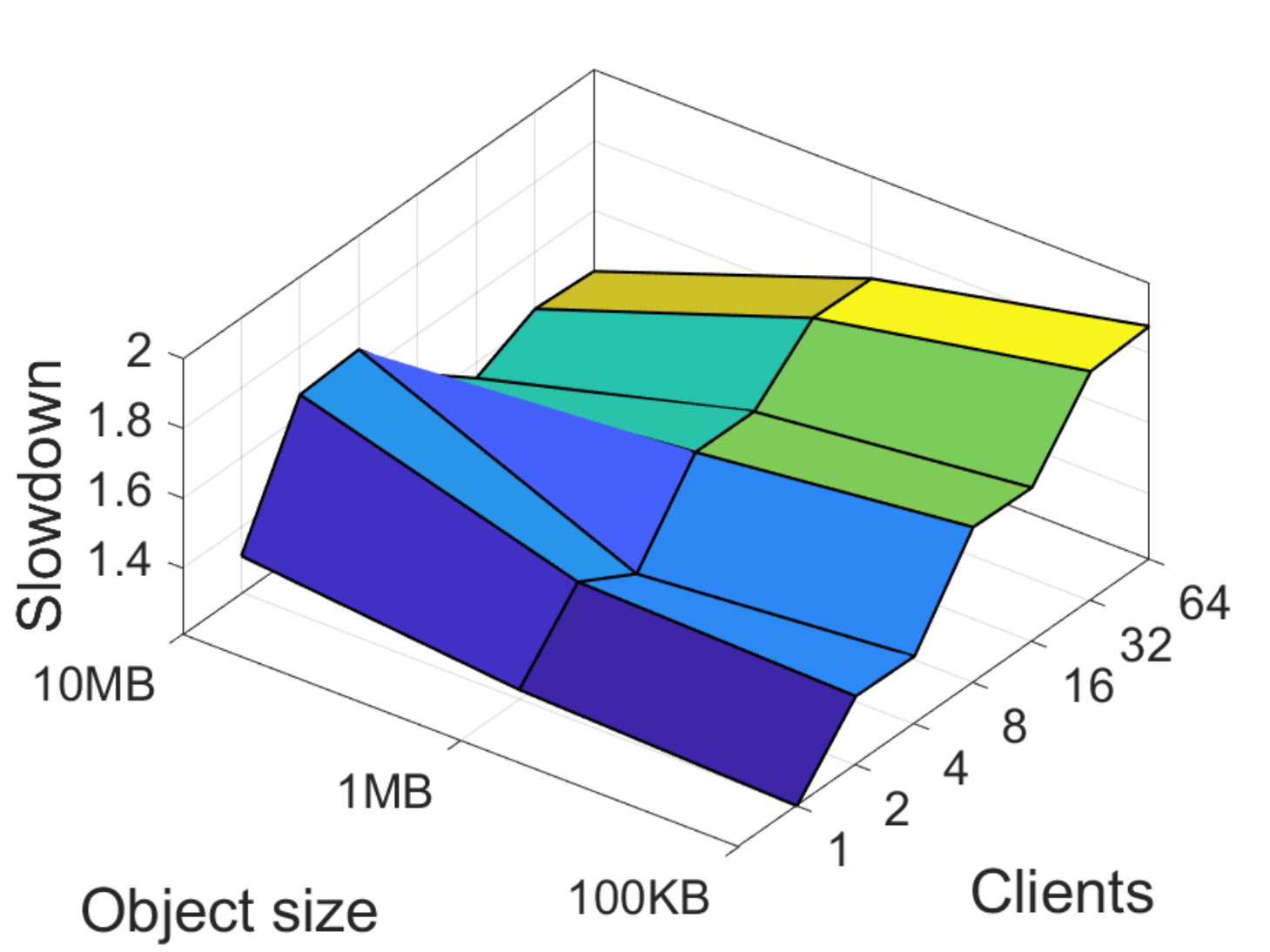}
\label{fig:4b}
}
\caption{Throughput and latency of \egeon~under a multi-client setting.}
\vspace{-15pt}
\end{figure*}

Also, Fig.~\ref{fig:4a} plots the throughput for an increasing number of emulated clients for a $1$MB object, which
exhibits the same behavior as before. That is, \egeon~and Zion showing a similar performance, while Swift
delivering a much higher throughput due to the absence of any computation in the I/O path. Finally, Fig.~\ref{fig:4b} illustrates 
\egeon's slowdown relative to vanilla Swift, calculated as $\mathrm{Slowdown = \frac{\egeon\,download\,time}{Swift\,download\,time}}$, 
as a function of the object size ($x$-axis) and the number of concurrent clients ($y$-axis). As can be seen in the figure, the slowdown
factor~does not increase steeply. Rather, it increases gradually in both axes, never doubling the latency. This indicates that pushing
down data protection logic to the storage may be acceptable for many applications.

\begin{table}[t!]
\centering
\caption{Maximum throughput (ops/sec) for different object sizes.}
\label{tab:IOPS}
\begin{tabular}{|c|c|c|c|}
\hline
\textbf{System} & \multicolumn{3}{c|}{\textbf{Object size}} \\ \cline{2-4}
& $100$KB & $1$MB & $10$MB \\ \hline\hline
Swift &  157,81 & 111,43 & 31,62\\ 
Zion &   140,72 & 79,69 & 21,17 \\ 
Egeon & 140,56 &79,59 & 21,26 \\  \hline
\end{tabular}
\vspace{-15pt}
\end{table}

\subsection{Applications}

To evaluate the composability of \egeon, we have designed two privacy policies that capture the 
different complexities of real-world applications. These applications are the following:

\smallskip
\noindent\textbf{Covid-$19$ use case.} In this use case, we demonstrate the same policy
of Listing~1, but applied to healthcare. We use the JSON file reported by the US government that
summarizes the patient impact on healthcare facilities caused by Covid-$19$~\cite{covid19} (April 2021).  
Specifically, we exchange the user label $\mathtt{``treasurer"}$ by  $\mathtt{``state\,coordinator"}$, and label the field
that reports the sum of patients hospitalized in a pediatric inpatient bed in~$7$-day periods [FAQ-$10$.b)] as
$\mathtt{``sensitive"}$, as it reveals which hospitals~may be collapsed. The rest of information is ignored.
As a result, the transformed view for state officials only bears a single homomorphically encrypted value
that aggregates the sum of all healthcare facilities. As in Listing~1, the policy links $3$ UDFs:
\texttt{CLAC}$\rightarrow$\texttt{HOM}$\rightarrow$\texttt{PRE}. To play out with the file size, we split
this dataset into three smaller files based on increasingly smaller time periods: year, month and week.

\smallskip
\noindent\textbf{Adult dataset~\cite{adult} use case.} This dataset in CSV format from
the UCI Machine Learning Repository~\cite{adult} has $48842$ records and $14$ attributes.
Some of these attributes can leak sensitive information such as race and occupation. For this
use case,~we have decided to protect the attribute $\#7$: occupation, with the \texttt{SEARCH}
scheme to allow for type-of-employment searches on encrypted text. To prove
composability, we have used \texttt{CLAC}~to protect three attributes out of the 
four attributes chosen for this experiment. We have used one object label: $\mathtt{``sensitive}"$, and  one user label:
 $\mathtt{``HR\,manager"}$, so that only a human resources manager can retrieve the three protected columns. Further,
we have encrypted other fields to increase the file size to $134$MB, to later split it into $3$ smaller files based on
the attribute  $\#6$: marital-status. The policy is as follows (some fields have been omitted for brevity):
\begin{lstlisting}[language=jsonnonumbersmall]
{ 
 "Object": "v1/{account}/{container}/adult.csv",
 "Action": {
  "StartAt": "Step1",
  "Steps": {
    "Step1": {
     "Id": "CLAC",
     "EventType": {
        "Type":  "ColumnMarkerEvent", 
        "Input": [{"columns": [2,6,7], "olabel": "sensitive" }]  
     },
     "Input": [{"ulabel": "HR manager", "olabel": "sensitive" }],
     "Next": "Step2" 
    },
    "Step2": {
     "Id": "SEARCH",
     "EventType": {"Type": "ColumnEvent", "Input": [{"column":7}]},
     "Next": "End"
 }}}
\end{lstlisting}

\begin{table}[h]
\caption{Network Speeds. }
\label{tab:NetworkSpeeds}
\centering
 \begin{tabular}{|l|l|l|l|}
 \hline
 \textbf{Network} & 4G & Fiber & LAN \\ \hline\hline
\textbf{Median speed (Mbps)}  &	$28.9$~\cite{4GUS}  & $55.98$~\cite{fiberUS} &	$887$ (SpeedTest) \\  \hline
\end{tabular}
\end{table}

\smallskip
\noindent\textbf{Experiment.} In this test, we measure the time to download~the raw files directly from Swift
against the time to download the transformed views generated by \egeon. The goal is to decide if it is worth 
to push the privacy transformations into the object store instead of running them on the user devices and VMs,~so that
software-defined data protection is within reach. To do so, we have capped the ingoing
bandwidth of our client machine to emulate different network speeds and customary scenarios: Fiber
network speeds to emulate home and business users, $4$G network bandwidth to emulate mobile users, and finally, LAN 
to simulate a scenario where the client and  Swift servers reside on the same local area network (e.g., a university
intranet). The exact network speeds are listed in Table~\ref{tab:NetworkSpeeds}.  We performed $1$K 
executions per object size and network speed.

\begin{figure}
\centering
\subfloat[][Covid-$19$ use case.]{
\includegraphics[width=0.245\textwidth]{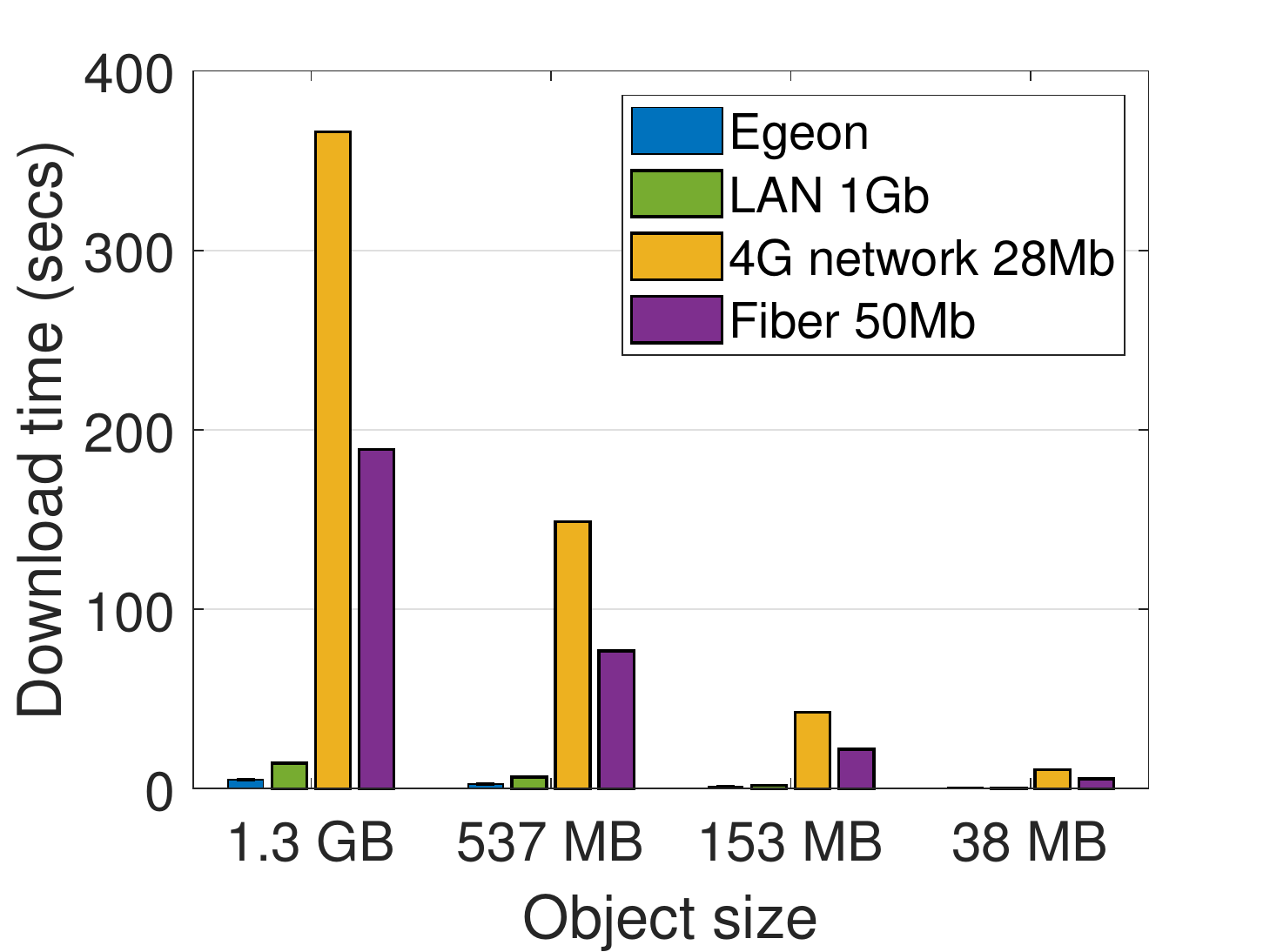}
	\label{fig:5}
}
\subfloat[][Adult dataset use case.]{
\includegraphics[width=0.245\textwidth]{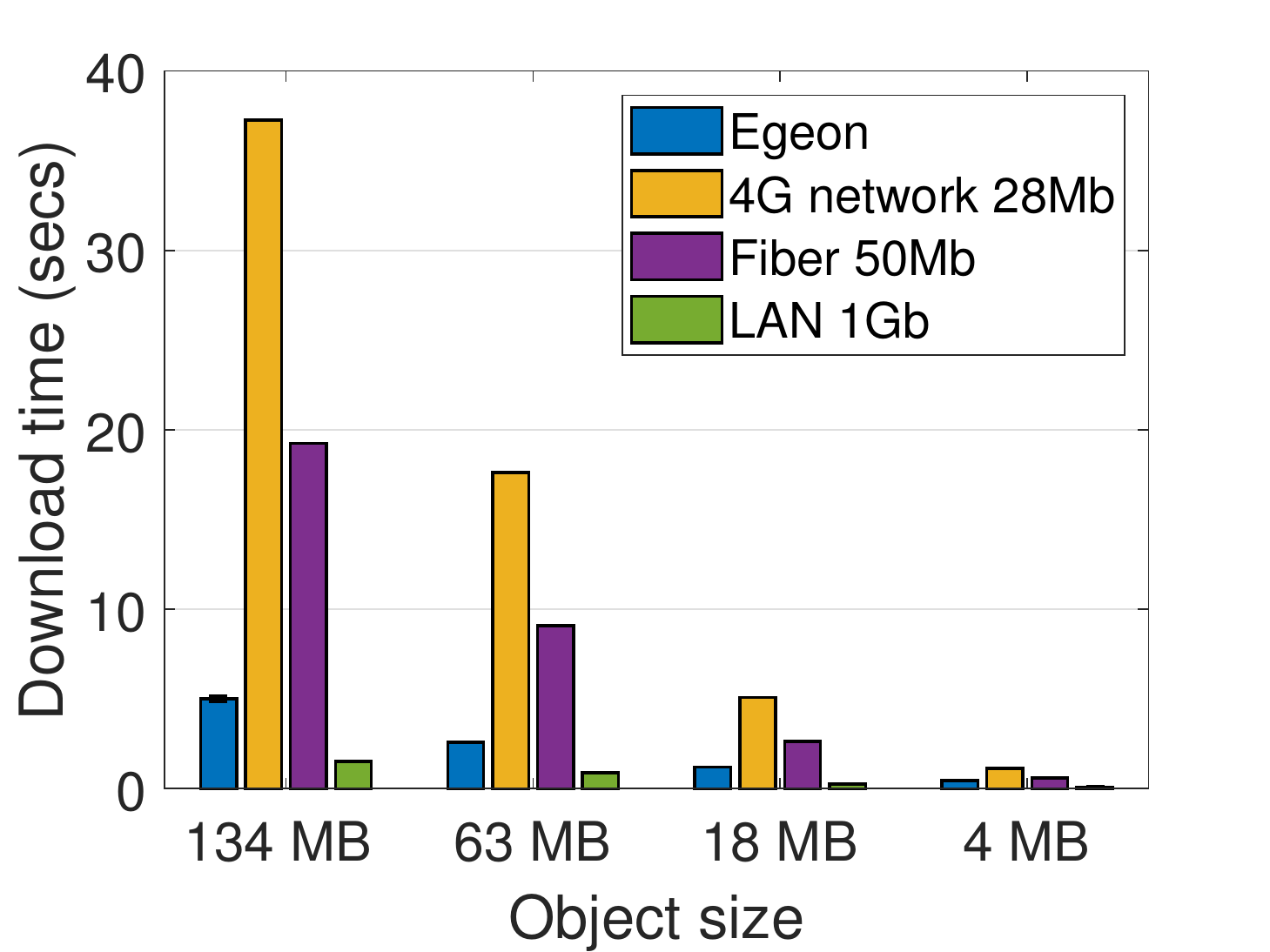}
\label{fig:6}
}
\caption{Performance of \egeon~in two composite policies.}
\vspace{-15pt}
\end{figure}

The results are plotted in Fig.~\ref{fig:5} for the Covid-$19$ use case, and in Fig.~\ref{fig:6} for the Adult dataset.
Error bars display the standard deviations of results, which are indeed very narrow. 
Non-surprisingly, we can see that \egeon~lowers the download time 
significantly for the slow $4$G and fiber connections, which means that pushing down privacy transformations
to storage~is a good deal better than the naive approach of encrypting data on the client side and retrieve
the whole file as alleged~by~cloud providers such as AWS for S$3$. The savings in some scenarios can be
dramatic such as in the Covid-$19$ use case, where just~a few bytes (e.g., aggregates such as \texttt{SUM}, \texttt{COUNT} and  \texttt{AVG}, etc.)
are consumed by the application, reaching $72.1$X speedup for $4$G mobile terminals. For the LAN setting, the benefits are not so
clear, and for the Adult dataset, \egeon~shows~a~slowdown factor of $3.3$X in the worst case due to the heavy
computations associated with keyword search---actually, the test primitive of \texttt{SEARCH} requires three pairing operations~\cite{peks}.
Either way,~we believe that \egeon's fine-grained data protection capabilities outweigh the slight loss of performance.

\section{Related Work}
 
\noindent\textbf{Software-defined storage systems}. A first category of related work comprises software-defined approaches for 
storage, and in particular, for object storage systems. The common feature of these approaches is that they break 
the vertical alignment of conventional storage infrastructures by reorganizing the I/O stack to decouple the control and data flows into two
planes~of functionality---control and data. A number of proposals have followed this approach, including IOFlow~\cite{ioflow}, sRoutes~\cite{sroute},  
Retro \cite{retro}, Vertigo~\cite{vertigo} and Crystal~\cite{crystal, FGCScrystal}. 

Among them, only Vertigo and Crystal have been tailored~to object storage. As \egeon, both systems have been deployed atop
OpenStack Swift. But unlike \egeon, their data plane~is based on OpenStack Storlets~\cite{storlet}. A Storlet is a piece of Java logic~that
is injected into the data plane to run custom storage services over incoming I/O requests. As in \egeon, this
design increases the modularity and programmability of the data plane stages, fostering reutilization. However, 
Storlet-made pipelines are not reactive, wasting resources when we are only interested in specific elements
of the data stream. Moreover, the Storlet-enabled data plane of Vertigo and Crystal emphasizes 
control-flow over data-flow, making it hard to explicitly represent the (cryptograhic) transformations of
objects. Per contra, \egeon's data plane is driven solely by the events showing up in the data streams, which
makes it easy to reuse the same transformations over and over again on different data. Only the events must~be
re-defined in the policies.

Finally, it is worth to note that to the best of our knowledge, we are not aware of another software-defined
storage~system that automatically enforces privacy policies along the I/O path as \egeon. Software-defined security 
has remained within the boundaries of software-defined networking (see, for instance, Fresco~\cite{fresco}).
The only exception is the recent vision paper~\cite{Zsolt21} on software-defined data protection. Like \egeon, 
\cite{Zsolt21} argues that the key ideas of software-defined storage can be translated to the data protection domain.
However, the approach of \cite{Zsolt21}~is radically different. Instead of adding privacy controls to  the I/O stack 
of a disk-based object storage system, \cite{Zsolt21} assumes all in-storage processing to occur on FPGAs and ``smart storage'' devices, 
for we see \cite{Zsolt21} as an orthogonal work to us.

\medskip
\noindent\textbf{Privacy Policy Enforcement}. There exist many systems that enforce
privacy policies automatically. Most of these systems resort to Information Flow Control (IFC) as a means to
control how information flows through the system. See, for instance, Riverbed~\cite{riverbed}, which
uses IFC to enforce user policies on how a web service should release sensitive user data. In contrast to these
systems, \egeon~follows a software-defined approach to leverage the storage resources and enforce the privacy
policies where data is. Similar to our transformation chains, Zeph~\cite{zeph} proposes to enforce privacy controls cryptographically
but over encrypted stream processing pipelines. Specifically for storage, Guardat~\cite{guardat}, at the block level, and Pesos~\cite{pesos}, at
the object level, enable users to specify security policies, for instance,~to
stipulate that accesses to a file require a record be added to an append-only log file. 
Nonetheless, these systems do not permit the composition of
advanced privacy controls as \egeon, and thus, fall short to empower users with
strong data controllers.

\section{Conclusions}

As increasingly much more sensitive data is being collected to
gain valuable insights, the need to natively integrate privacy
controls into the storage systems is growing in importance. In
particular, the poor interface of object storage systems, 
which lacks of sophisticated data protection mechanisms, along with
the inherent difficulties to refactor them, have motivated us~to
design and implement \egeon.  To put in a nutshell, \egeon~is a novel software-defined data protection framework for object storage. 
It allows data owners to define privacy policies on~how their data can be shared, which permit the composition of 
data transformations to build sophisticated data protection controls. In this way, data owners can specify multiple views from the same data piece, 
and modify these views by only updating the policies (e.g., by modifying the chain of transformations that produce a particular view), leaving the
system internals intact.  The  \egeon\hphantom{} prototype has  been  coded atop
OpenStack Swift. And our evaluation results demonstrate that \egeon\hphantom{} adds little 
overhead to the system, yet empowering users with~the needed controls to ensure strong data protection. 

\section*{Acknowledgment}

This research has been partly supported
by EU under grant agreement No. 825184 and the Spanish
Government through project PID2019-106774RB-C22. Marc
Sanchez-Artigas is a Serra-H\'{u}nter Fellow.

\bibliographystyle{IEEEtran}
\bibliography{References}

\end{document}